\documentclass[ reprint,floatfix,aps,]{revtex4-1}

\usepackage{graphicx,amsmath,amssymb,amsfonts}

\usepackage[caption=false]{subfig}
\usepackage{dcolumn}

\begin{document}

\preprint{APS/123-QED}

\title{Scaling Fixed-Field Alternating-Gradient accelerators with reverse bend and spiral edge angle}
\author{Shinji Machida}
 \email{shinji.machida@stfc.ac.uk}
\affiliation{ STFC Rutherford Appleton Laboratory \\
 Harwell Campus, Didcot, OX11 0QX, United Kingdom}


\begin{abstract}
A novel scaling type of Fixed-Field Alternating-Gradient (FFAG) accelerator is proposed that solves the major problems of conventional scaling FFAGs. This scaling FFAG accelerator combines reverse bending magnets of the radial sector type and a spiral edge angle of the spiral sector type to ensure sufficient vertical focusing without relying on extreme values of either parameter. This new concept makes it possible to design a scaling FFAG for high energy (above GeV range) applications such as a proton driver for a spallation neutron source and an accelerator driven subcritical reactor.
\end{abstract}

\maketitle

Particle accelerators were developed initially as a tool to explore particle physics at the energy frontier. Recently, however, many accelerators have been constructed for other fields of physics mostly with the aim of producing secondary or tertiary particles such as neutrons, muons and neutrinos. A figure of merit in this area is a measure of the number of energetic particles, usually protons, that are used to create secondary or tertiary particles through impact with a production target. The energy of each particle does not have to be as high as in accelerators for research at the energy frontier; instead emphasis is put on the beam intensity, which is always demanding. The research field that this type of accelerator explores is called the intensity frontier, and the accelerator is usually referred to as a proton driver.

Considering the cross-section of the secondary and tertiary particle production, the energy range of a proton driver covers a range from a few 100\,MeV to some 10's of GeV. Cyclotrons cover the lower end: the machines at PSI and TRIUMF, for example, have just enough energy for neutron and muon production. Linear accelerators (linacs for short) with and without an accumulator ring and rapid cycling synchrotrons (RCS) usually produce beams of a few GeV to produce neutrons most efficiently. ISIS, SNS, the J-Parc RCS and ESS (under construction in Sweden) belong to this category. When protons with energies higher than a few GeV are required for the production of kaons and neutrinos through pions, slow cycling synchrotrons are the only option. BNL-AGS, CERN-PS and J-Parc MR are examples.

Fixed-Field Alternating-Gradient (FFAG) accelerators were invented in 1950s and developed over the following years, initially as accelerators for energy frontier physics \cite{Symon, Kolomensky}. At the same time, an alternating-gradient synchrotron had been developed and its more compact magnets relative to the FFAGs became a big advantage when looking to increase beam energy, so the objectives of the FFAG accelerator development faded out. Although there were remained pockets of interests on FFAG accelerators, for instance \cite{Ishikawa, Kustom, Meads, Martin, Jungwirth}, little development beyond paper studies took place until the late 1990's when the idea of a neutrino factory called for an accelerator that could rapidly accelerate muons before they had time to decay \cite{Mills, Johnstone1, Machida1}.

When FFAGs were invented, it was realized that an important advantage over other type of accelerator is its potential for high beam intensity with an energy range covering a few GeV. Although CW operation of cyclotrons is the simplest way to obtain high average beam intensity, the energy range is limited below  $\sim$1\,GeV. At higher energies, its size becomes too large and also beam extraction becomes difficult because the turn separation at the outer orbits is minimal. Although the fixed field nature of FFAGs requires relatively large magnets to cover the orbit excursion from injection to extraction energy, the fixed field nature also enables rapid acceleration as well as high repetition rate of operation as long as the rf acceleration system can provide sufficient power. This combines to satisfy the needs of intensity frontier accelerators.

In the last 15 years, there has been significant progress in the development of FFAG accelerators. For high intensity applications, a proof of principle model with 1\,MeV output energy was constructed at KEK~\cite{Aiba}. Two scaled-up machines, one a prototype for medical applications~\cite{Adachi} and the other for a proton driver to drive a sub-critical reactor (ADSR)~\cite{Tanigaki} were constructed at KEK and Kyoto University, respectively.

Both machines follow the scaling FFAG design and have a vertical magnetic field profile given by
\begin{equation}
B_z\left(r,\theta\right)=B_{z0}\left( \frac{r}{r_0} \right)^{k} F\left( \vartheta \right), \label{eq:one}
\end{equation}
where
\[
\vartheta = \theta-\tan\delta\,\ln\frac{r}{r_0}, \label{eq:two}
\]
is the generalized azimuthal angle, $r$ is the radial coordinate, $\theta$ is the geometrical azimuthal angle, $r_0$ and $B_{z0}$ are the reference radius and the vertical magnetic field at the reference radius, respectively. $k$ is the geometrical field index defined as
\[k=\frac{r}{B_z}\left( \frac{\partial B_z}{\partial r} \right). \]
$F(\vartheta)$ is a periodic function with period $2\pi/N$, where $N$ is the number of cells in the ring. $\delta$ is the spiral edge angle.

With this magnetic field profile, the scaling FFAG satisfies the scaling conditions,
\begin{align}
\left.\frac{\partial}{\partial p}\left(\frac{K}{K_0}\right)\right|_{\vartheta=const.}&=0, \label{eq:three}\\
\left.\frac{\partial k}{\partial p}\right|_{\vartheta=const.}&=0, \label{eq:four}
\end{align}
where $K$ is the curvature of the orbits and $K_0$ refers to its reference orbit, $p$ is the beam momentum.
The scaling conditions make the transverse tune of strong focusing accelerators constant with fixed field magnets and avoid resonance crossing during acceleration. For extremely fast acceleration for short lived particles like muons, however, this can be violated, which leads to the concept of a non-scaling FFAG~\cite{Johnstone2}. Very fast acceleration without the scaling condition was first demonstrated in the EMMA project in the U.K. in 2012~\cite{Machida2}.

In practice, scaling FFAGs are realized by two different types of structure. One is based on radial sector magnets~\cite{Cole} and the other uses a spiral sector structure~\cite{Kerst} and depends on the form of
$F(\vartheta)$ in Eq.~\eqref{eq:one}.
A radial sector FFAG employs the function $F(\vartheta)$ to flip the sign periodically so that normal and reverse bending magnets provide alternating focusing. In a spiral sector FFAG, the function $F(\vartheta)$ is always positive with only normal bending magnets, but the magnet pole face has a finite edge angle with respect to the orbits, which gives the lattice magnets a spiral shape when viewed from above. The proper edge angle introduces a strong defocusing in the horizontal direction as opposed to the focusing in the body field of the magnets. Both the radial and spiral sector FFAGs that were constructed in the 1950's accelerated electrons to a few 100\,keV. Recently a spiral FFAG for proton acceleration up to 2.5\,MeV and two radial FFAGs up to 150\,MeV were constructed in Japan~\cite{Tanigaki}.

In order to go beyond the prototype machines and be competitive with linacs and synchrotrons in energies beyond a GeV, FFAGs faced practical problems. The number of cells has to increase to keep the individual magnets within reasonable field strengths and lengths. As a result, the bending angle per cell becomes relatively small. Either the spiral angle should be large or the strength of the reverse bending magnets should be high to keep enough vertical focusing.
This problem was not seen in the prototype FFAGs which do not have as many cells because of their lower energies.

This paper proposes a novel scaling FFAG that solves the difficulties by combining the principles of radial and spiral FFAGs together. It is referred to as the \emph{DF-spiral FFAG} and can be regarded as either a small spiral angle addded to a radial FFAG or a small reverse bend added to a spiral FFAG.

In scaling FFAGs, the ring tunes $Q_{x,z}$ are approximated by the following equations, \eqref{eq:five} and \eqref{eq:six}, as long as $Q_{x,z}^2\ll (N/2)^2$~\cite{Kolomensky}, as
\begin{align}
Q_x^2 & \approx 1-k+\dfrac{k^2S^2}{N^2b_0^2}, \label{eq:five}\\
Q_z^2 & \approx k+\dfrac{k^2S^2}{N^2b_0^2}+\dfrac{\Phi^2}{b_0^2}\left( 1+2\tan^2\delta \right), \label{eq:six}
\end{align}
where the $\{b_k\}$ are defined as Fourier expansion coefficients for the vertical field in the azimuthal direction:
\[B_z =B_{z0}\sum_{k=0}^\infty b_k e^{ikN\theta}\]
and
\begin{align*}
\Phi^2 & =4\sum_{k=1}^\infty |b_k|^2,\\
S^2 & =2\sum_{k=1}^\infty \dfrac{|b_k|^2}{k^2},
\end{align*}
where $N$ is the total number of cells in the ring lattice. The quantity $\Phi/b_0$ is called the field flutter. The term $2\Phi^2\tan^2\delta/b_0^2$ is a measure of the specific strong focusing due to the spiral field shape.
In short, the vertical tune is a function of the field flutter $\frac{\Phi}{b_0}$ and the spiral angle $\delta$. In a radial sector FFAG, the spiral angle is zero and the tune is dominated by the field flutter. In a spiral sector FFAG, the tune is adjusted by the spiral angle because the field flutter is almost unity since there are only normal bending magnets.

\begin{figure}[b]
\includegraphics[bb=0 0 365 200,width=.95\columnwidth]{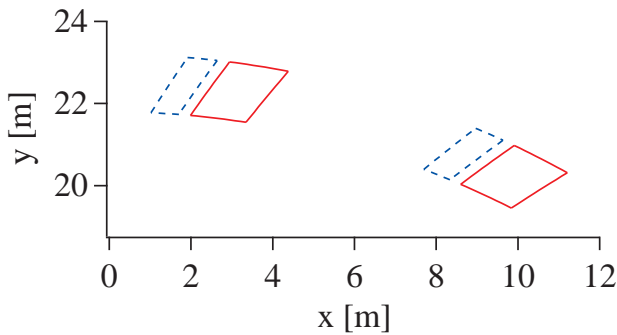}
\caption{\label{fig:lattice} Top view of DF-spiral FFAG lattice. The red solid line indicates a normal bending magnet
and the blue dashed line identifies a magnet with reverse bend. Coordinates (0,0) give the machine centre and the
orbit radius is about 23\,m. The spiral angle is $30^\circ$ in this example. Refer to Table~\ref{tab:mainparameters}
for other parameters.} \end{figure}

It is clear that making both the field flutter and the spiral angle adjustable at the same time gives more flexibility and better optimisation for the vertical focusing without relying on extreme values of either parameter.
This can be realized, for instance, by placing normal and reverse bending magnets next to each other with a finite
spiral angle to make a doublet focusing cell as shown in Fig.~\ref{fig:lattice}. As a result, the edge focusing is
enhanced in the vertical direction as we show below.

In order to obtain a more quantitative estimate, consider a design of a 1.2\,GeV proton machine as an example. It
consists of 20 identical cells with a $3.6^\circ$ normal bending magnet (Bf) and a $1.8^\circ$ reverse bending magnet
(Bd). The ratio of integrated Bd and Bf (absolute) strengths, the spiral angle and the geometrical field index are
the three, free parameters that we explore. The nominal average radius is 23\,m so the maximum field strength is
within rthe reach ($\sim$1.8\,T) of normal conducting magnets. The long drift space is about 5\,m, which is enough
for the injection and extraction systems and the rfcavities. The main parameters are listed in
Table~\ref{tab:mainparameters}.

\begin{table}[b]
\caption{\label{tab:mainparameters}
Main parameters of the test lattice.}
\begin{ruledtabular}
\begin{tabular}{lcc}
Parameter & Value & Unit\\
\colrule
Number of cell & 20 & - \\
Nominal radius & 23 & m\\
Effective length of Bd & 1.8 & degs\\
Fringe length of Bd & 0.75 (Bf side) & degs\\
 & 1.5 (other end) & degs\\
Effective length of Bf & 3.6 & degs\\
Fringe length of Bf & 0.75 (Bd side) & degs\\
 & 1.5 (other end) & degs\\
Short drift Space \\ between Bd and Bf & 0.75 & degs\\
Long drift space & 10.35 & degs
\end{tabular}
\end{ruledtabular}
\end{table}

The edges of the Bd and Bf magnets are curved with a non-zero spiral angle and the field falls off as a half sinusoidal curve on both side. To ensure the scaling conditions, the vertical magnetic field on the mid-plane is modelled as a function of the azimuthal angle as
\begin{alignat*}{2}
B&_z(r,\theta,0)=\\
&\left(1+\sin\left( \pi\frac{\left( \theta-\theta_{b1}-\tan\delta\ln\left( r/r_0 \right) \right)}{\Delta\theta_{f1}} \right) \right) B_{z0}\left( \dfrac{r}{r_0} \right)^k
\\
 &\text{for}\qquad\theta_{b1}-\Delta\theta_{f1}/2+\tan\delta\ln\left(r/r_0\right)<\theta \\
&\qquad\qquad\qquad<\theta_{b1}+\Delta\theta_{f1}/2+\tan\delta\ln\left(r/r_0\right);
\\
  B&_z(r,\theta,0)=B_{z0}\left( \dfrac{r}{r_0} \right)^{k}\\
 &\text{for}\qquad\theta_{b1}+\Delta\theta_{f1}/2+\tan\delta\ln\left(r/r_0\right)<\theta \\
 &\qquad\qquad\qquad<\theta_{b2}-\Delta\theta_{f2}/2+\tan\delta\ln\left(r/r_0\right);\\
  B&_z(r,\theta,0)=\\
  &\left(1+\sin\left(\pi\frac{\left(-\theta+\theta_{b2}+\tan\delta\ln\left(r/r_0\right) \right)}{\Delta\theta_{f2}}\right)\right)B_{z0}\left(\dfrac{r}{r_0} \right)^k\\
 &\text{for}\qquad\theta_{b2}-\Delta\theta_{f2}/2+\tan\delta\ln\left( r/r_0 \right)<\theta  \\
 &\qquad\qquad\qquad<\theta_{b2}+\Delta\theta_{f2}/2+\tan\delta\ln\left( r/r_0 \right);
\end{alignat*}
where $\theta_{b1}$ and $\theta_{b2}$ are the azimuthal positions of the effective boundaries, $\Delta\theta_{f1}$ and $\Delta\theta_{f2}$ are the lengths of the fringe regions. The magnetic fields in other directions ($B_r$ and $B_\theta$) as well as $B_z$ off the mid-plane are derived from Maxwell's equations up to the sixth order in $z$.

Once the lattice magnets are specified, the multi-particle tracking code Scode~\cite{Machida3} is used to
calculate the ring optics and the particle beam dynamics. The equilibrium orbits for different momenta are
found iteratively. A one-turn (or one-cell) transfer map is constructed using several test particles with
different initial conditions with small amplitudes in each coordinate. The betatron tunes and lattice
functions are calculated based on this map.

\begin{figure}[htb]
  \subfloat[vertical\label{fig:k17_sub1}]{%
  \includegraphics[bb=0 0 370 350,width=.45\columnwidth]{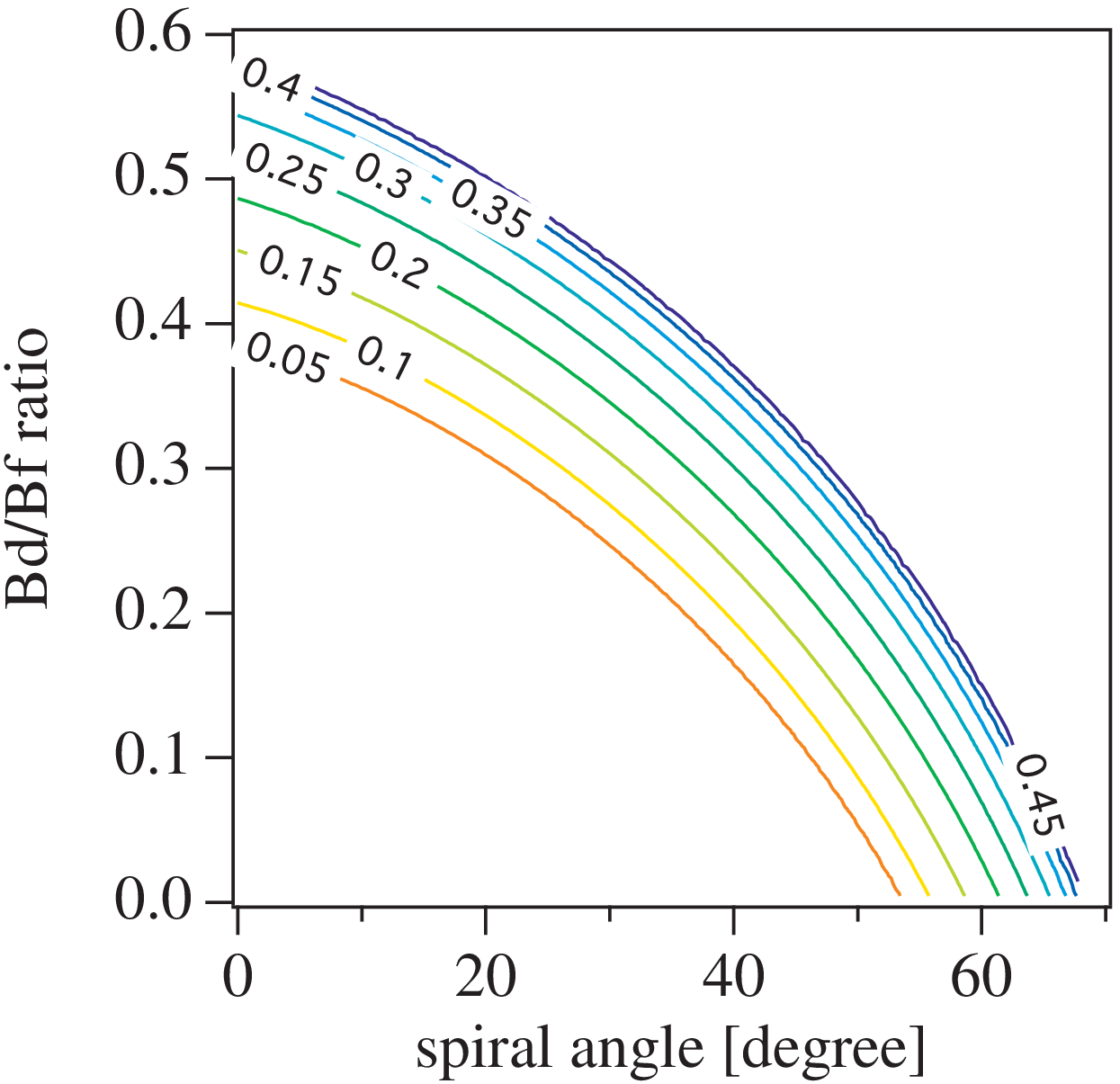}
  }
\hfill
  \subfloat[horizontal\label{fig:k17_sub2}]{%
  \includegraphics[bb= 0 0 370 350,width=.45\columnwidth]{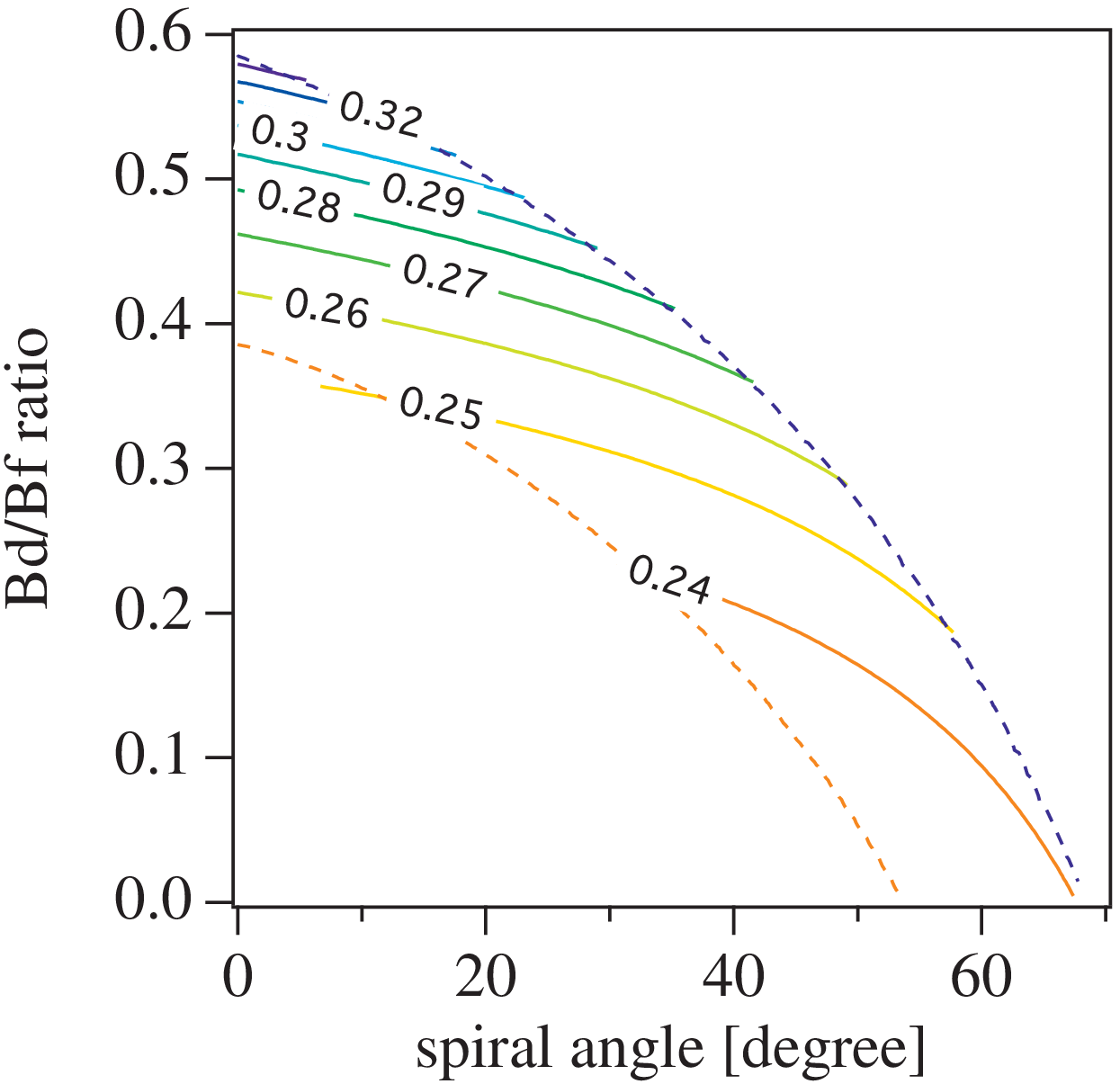}
  }
\caption{Stable area with cell tune indicated as contour curve when $k$=17.}
\label{fig:k17}
\end{figure}

The advantage of the DF-spiral configuration is illustrated in Fig.~\ref{fig:k17} where an optical study is
made using the ratio of integrated Bd and Bf strength to represent the field flutter, as in
Eq.~\eqref{eq:six}, for varying spiral angle. The geometrical field index is fixed at $k=17$.

Figure.~\ref{fig:k17_sub1} shows the domain that gives stable betatron oscillations. The contours
correspond to vertical cell tunes from 0 to 0.5 ($0^\circ$ to $180^\circ$ phase advance). It should be
noted that in a conventional radial sector FFAG, the variable parameter is the Bd/Bf ratio and is allowed to move only on the $y$-axis.
When the phase advance per cell is around $90^\circ$, the Bd/Bf ratio has to be around 0.5, which makes
the machine circumference very large. In a conventional spiral sector FFAG, on the other hand, the spiral angle is the variable
parameter and so the variation is along the $x$-axis. A spiral angle of about $60^\circ$ is not entirely
impractical, but the main lattice magnets becomes very complex. Now we have the whole 2D region in
parameter-space which gives us moderate choices for the spiral angle and the flutter factor
simultaneously. Figure~\ref{fig:k17_sub2} shows the same stable regions for varying horizontal cell tune.
The horizontal cell tune is less sensitive to the parameters although it is clear that the higher Bd/Bf
ratio leads to higher horizontal tune. The whole stable area is determined by the vertical stability.

\begin{figure}[htb]
  \subfloat[vertical\label{fig:k25_sub1}]{%
  \includegraphics[bb=0 0 370 350,width=.45\columnwidth]{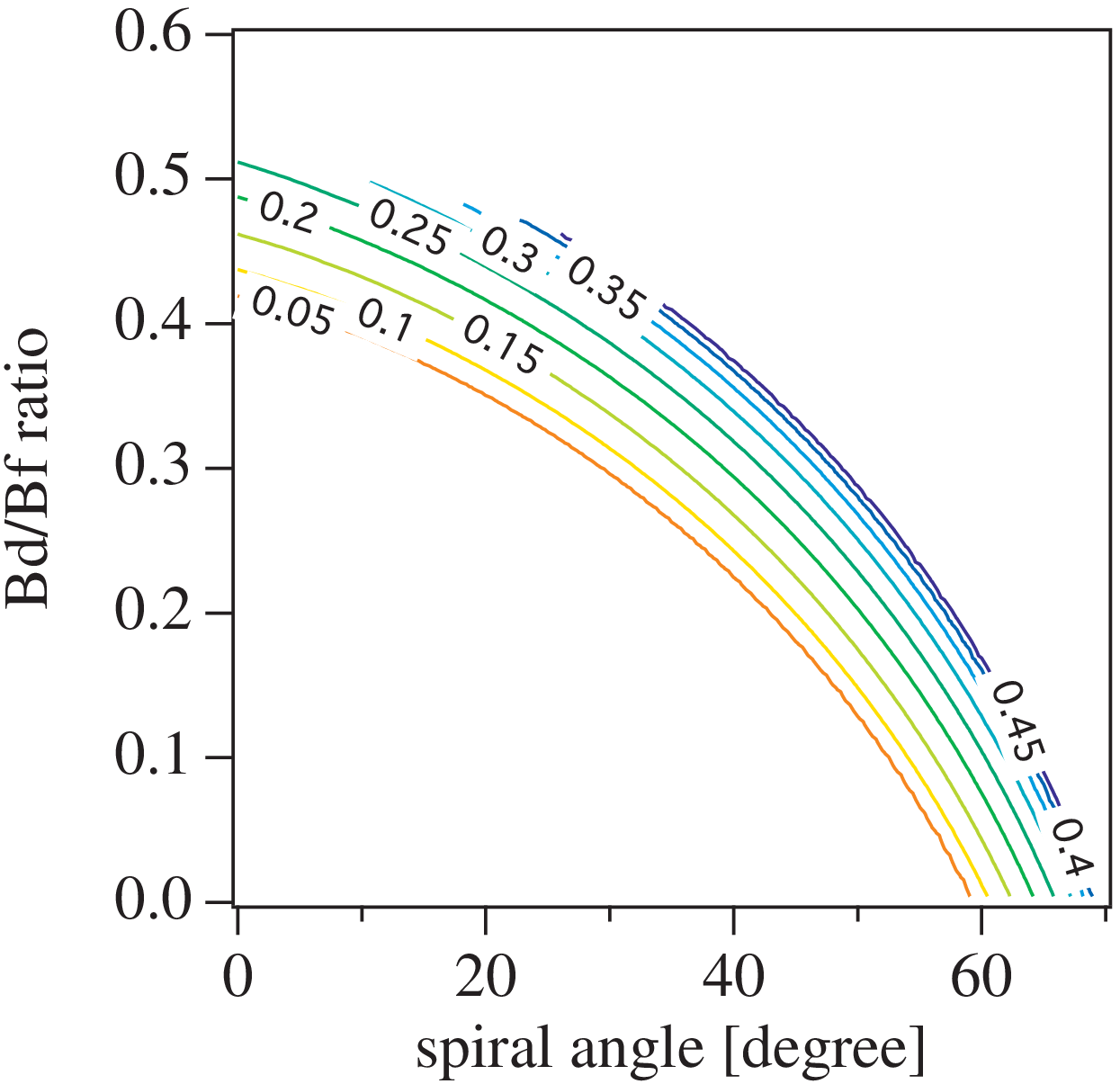}
  }
\hfill
  \subfloat[horizontal\label{fig:k25_sub2}]{%
  \includegraphics[bb= 0 0 370 350,width=.45\columnwidth]{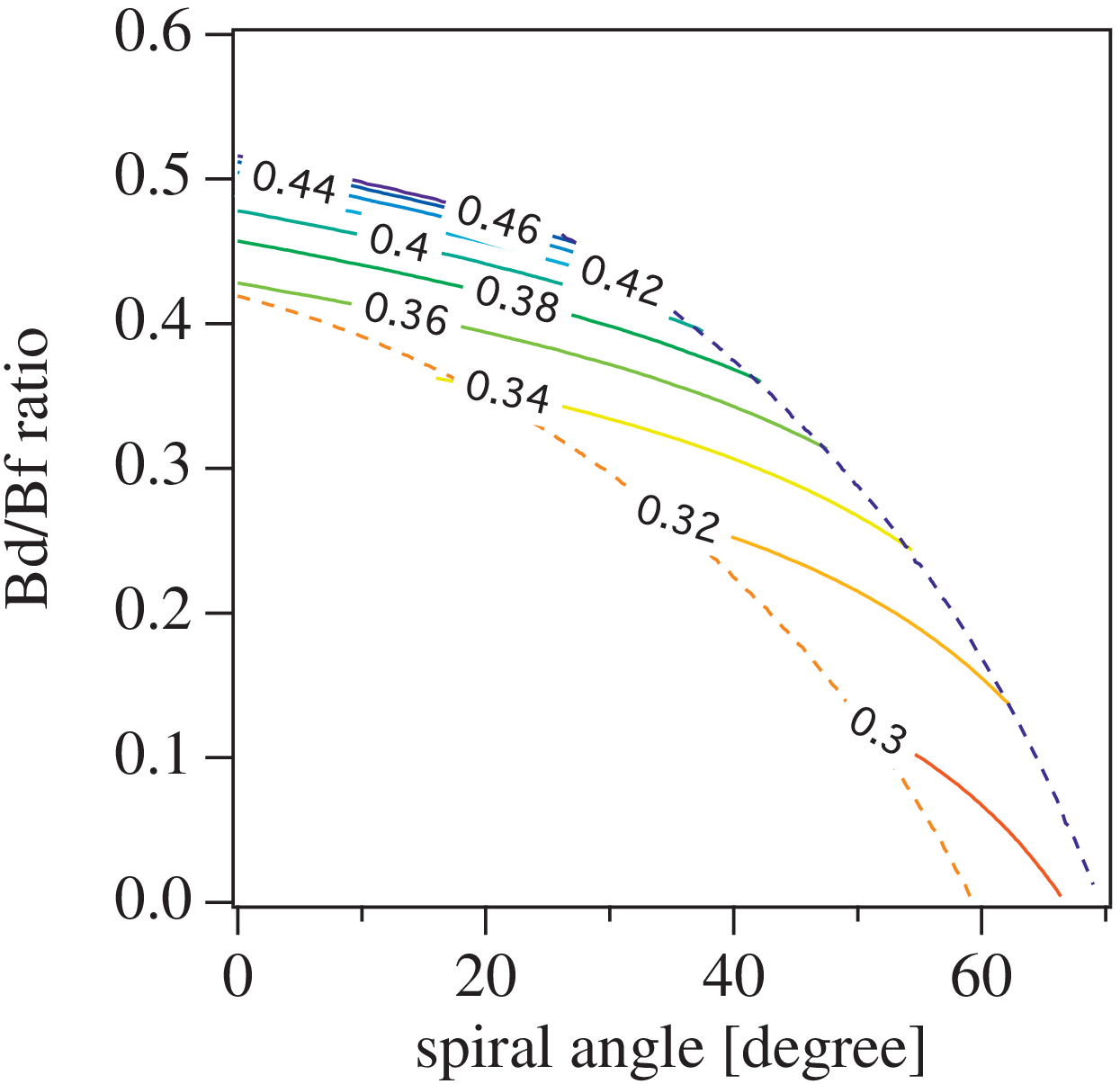}
  }
\caption{Stable area with cell tune indicated as a contour curve when $k$=25.}
\label{fig:k25}
\end{figure}

With larger value of $k$,  e.g. $k=25$, the stable area shrinks as shown in Fig.~\ref{fig:k25_sub1}. The variation of horizontal cell tune becomes larger within the stable area as shown in Fig.~\ref{fig:k25_sub2}.
The whole stable area is still mainly determined by the vertical cell tune, but in the region of high Bd/Bf, the horizontal cell tune reaches 0.5 and sets the stability boundary.

One of the major concerns of FFAG accelerators is dynamic aperture, which may become deteriorated by intrinsic nonlinearities of the lattice magnets~\cite{Yoshimoto}.
The DF-spiral concept is not an exception.
Two different constraints are imposed in order to explore dynamic aperture in the tune space. The first fixes the spiral angle at $30^\circ$ and adjusts the Bd/Bf ratio together with $k$. The other fixes the Bd/Bf ratio at 0.23 and adjusts the spiral angle together with $k$. We label the former ``DF-spiral A" and the latter ``DF-spiral B".

The absolute strength of the magnets was adjusted to make the average orbit around 23\,m for 1.2\,GeV proton beams. The dynamic aperture is defined as the initial horizontal amplitude with which a particle can survive for 10000 turns at a fixed energy of 0.4\,GeV, which is the nominal injection energy of the 1.2\,GeV FFAG accelerator. The 10000 turns corresponds to a time scale of 10\,ms in this size accelerator. The whole acceleration period is supposed to finish within this time period. Synchrotron oscillations are ignored. The initial vertical amplitude was fixed at 100\,$\pi$,mm.mrad.

\begin{figure}[htb]
  \subfloat[DF-spiral A\label{fig:DF_sub1}]{%
  \includegraphics[bb=100 55 360 280,width=.48\columnwidth]{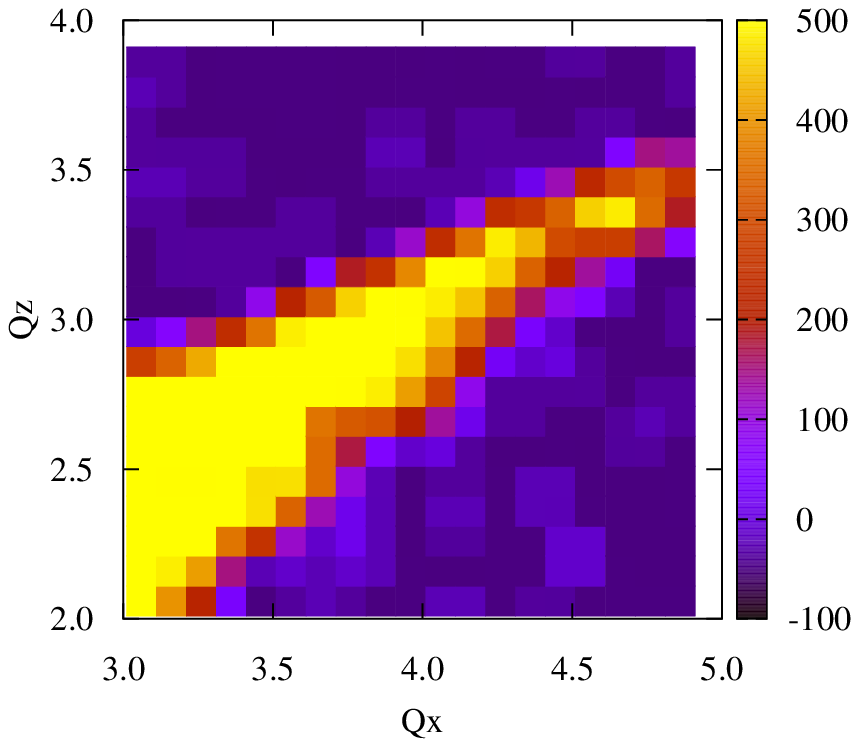}
  }
\hfill
  \subfloat[DF-spiral B\label{fig:DF_sub2}]{%
  \includegraphics[bb= 100 55 360 280,width=.48\columnwidth]{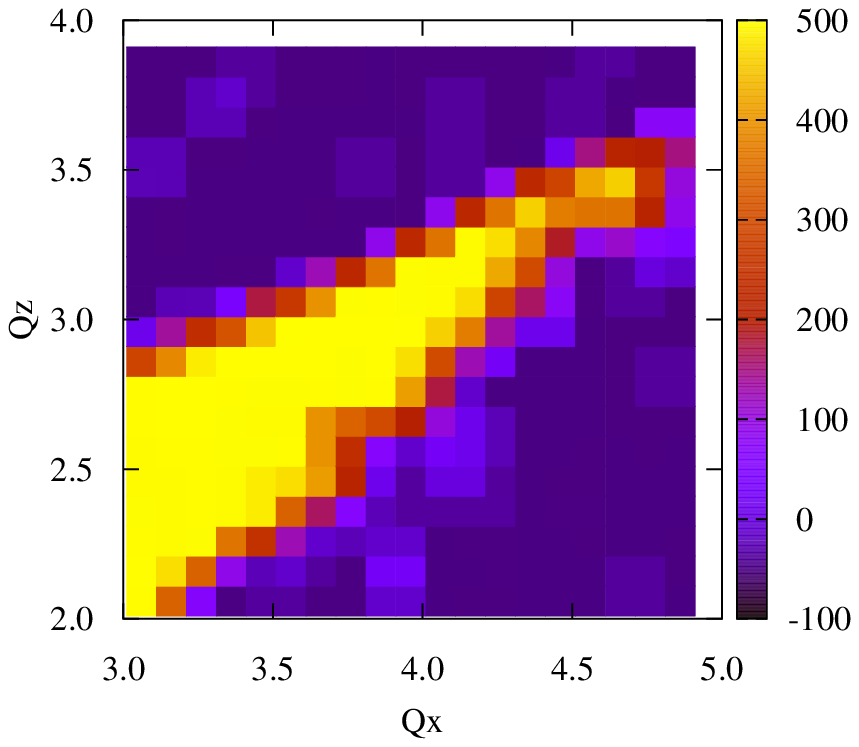}
  }\\
  \subfloat[radial sector\label{fig:DF_sub3}]{%
  \includegraphics[bb=100 55 360 280,width=.48\columnwidth]{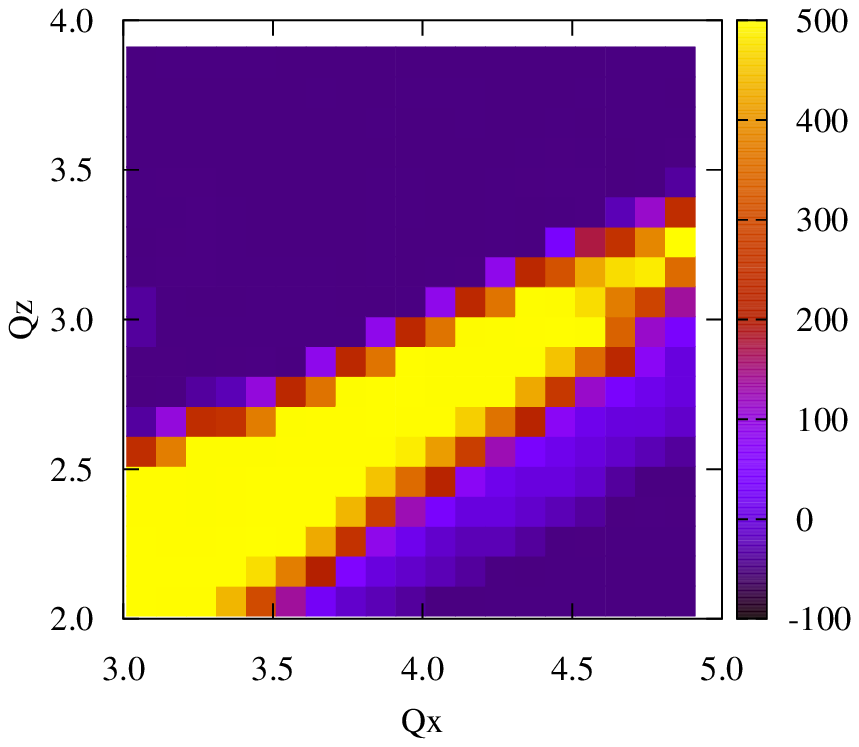}
  }
\hfill
  \subfloat[spiral sector\label{fig:DF_sub4}]{%
  \includegraphics[bb=100 55 360 280,width=.48\columnwidth]{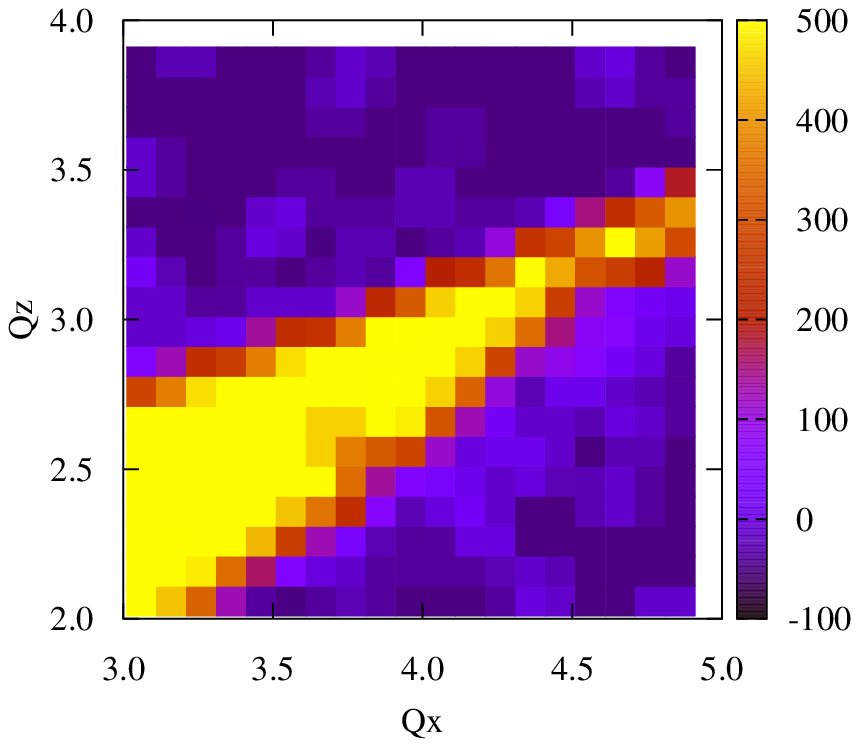}
  }
\caption{ Dynamic aperture of four lattice configurations in cell tune space $Q_x$=3.0 to 5.0 and $Q_z$=2.0
to 4.0. Dynamic aperture is defined as the maximum initial horizontal amplitude leading to particle
survival for 10000 turns. The initial vertical amplitude is 100\,$\pi$\,mm.mrad. The colour scale refers to
horizontal amplitude in units of $\pi$\,mm.mrad.}
\label{fig:DA}
\end{figure}

In both DF-spiral A and B, horizontal dynamic apertures of more than 500\,$\pi$\,mm.mrad are achieved at a certain tune region as shown in Figs.~\ref{fig:DF_sub1} and \ref{fig:DF_sub2}, respectively. Note that 500\,$\pi$\,mm.mrad is the maximum aperture we have explored and more than the physical aperture of similar energy proton drivers under operation~\cite{Cousineau, Hotchi}. The dynamic aperture in conventional radial and spiral FFAGs in Figs.~\ref{fig:DF_sub3} and \ref{fig:DF_sub4} shows almost no difference between the different lattice configurations under study.

A crucial question in the design fixed field accelerators, both cyclotrons and FFAGs, is how to ensure
sufficient vertical focusing. In particular, FFAGs need strong focusing comparable with the focusing in
horizontal direction. Employing strong reverse bending magnets tends to enlarge the average machine radius
in a radial sector FFAG. Employing large spiral edge angles makes spiral FFAG magnets practically
unachievable. This was not regarded as a crucial issue for moderate maximum energies, up to a few 100\,MeV,
say. The issue becomes more pronounced when a higher energy FFAG above 1\,GeV is needed where the number of
cells increases significantly, usually much more than 10. The increase in radius can have a significant
impact, especially if the machine radius is already large. We need significantly larger edge angles when the
bending angle per cell is small and almost no vertical focusing is provided from the edge.

In this paper, a novel scaling FFAG has been proposed which has features of both conventional radial and spiral sector FFAGs. The name ``DF-spiral FFAG" is suggested. Having simultaneous vertical strong focusing from reverse bending magnets and spiral edge focusing eases the requirement from each function and provides increased confidence that such a design is achievable.

A 1.2\,GeV proton FFAG design with a 20 cell lattice has been used as an example. The important property of dynamic aperture has been calculated and shows no reduction in aperture compared with conventional radial and spiral sector FFAGs. The study demonstrates that there are indeed advantages in the DF-spiral design, which could well play a part in the development of future fixed field accelerators.

We wish to acknowledge the encouragement by Christopher Prior and support from members of the ISIS department at Rutherford Appleton Laboratory.

\end{document}